\documentclass{aa}
\usepackage{graphicx}
\usepackage{natbib}
\usepackage{epsfig}
\def\ut#1{\mathop{\vtop{\ialign{##\crcr
     $\hfil\displaystyle{#1}\hfil$\crcr\noalign
     {\kern1pt\nointerlineskip}\hbox{$\hfil\sim\hfil$}\crcr
     \noalign{\kern1pt}}}}}

\def\undersymbol#1#2{\mathop{\vtop{\ialign{##\crcr
     $\hfil\displaystyle{#2}\hfil$\crcr\noalign
     {\kern1pt\nointerlineskip}\hbox{$\hfil#1\hfil$}\crcr
     \noalign{\kern1pt}}}}}
\def\arcsec{^{\prime\prime}}

\begin{document}

 \title{$XMM$-Newton and Swift observations of WZ Sge: spectral and timing analysis}
\author{A.A. Nucita\inst{1,2}, E. Kuulkers\inst{3}, F. De Paolis\inst{1,2}, K. Mukai\inst{4,5}, G. Ingrosso\inst{1,2}, B.M.T. Maiolo\inst{1,2}}
\institute{Department of Mathematics and Physics {\it E. De Giorgi}, University of Salento, Via per Arnesano, CP 193, I-73100,
Lecce, Italy \and INFN, Sez. di Lecce, via per Arnesano, CP 193, I-73100, Lecce, Italy \and European Space Astronomy Centre, SRE-O, P.O.~Box 78, 28691,
Villanueva de la Ca\~nada (Madrid), Spain \and CRESST and X-ray Astrophysics Laboratory, NASA Goddard Space Flight Center, Greenbelt, MD 20771, USA \and Department of
Physics, University of Maryland Baltimore County, 1000 Hilltop Circle, Baltimore, MD 21250, USA}

\offprints{A.A. Nucita, \email{nucita@le.infn.it}}
\date{Submitted: XXX; Accepted: XXX}

{
  \abstract
   {WZ Sagittae is the prototype object of a subclass of dwarf novae, with rare and long (super)outbursts, in which a white dwarf primary accretes matter from
    a low mass companion. High-energy observations offer the possibility of
    a better understanding of the disk-accretion mechanism in WZ Sge-like binaries.}
    {{ We used archival {\it XMM}-Newton and Swift data to characterize
     the X-ray spectral and temporal properties of WZ Sge in quiescence.}}
    {{ We performed a detailed timing analysis of the simultaneous X-ray and UV
     light curves obtained with the EPIC and OM instruments
     on board {\it XMM}-Newton in 2003.  We employed several techniques in
     this study, including a correlation study between the two curves.
     We also performed an X-ray spectral analysis using the EPIC data, as well
     as Swift/XRT data obtained in 2011.}}
   {We find that the X-ray intensity is clearly modulated at a period of 
    $\simeq 28.96$ s, confirming previously published preliminary results.
    We find that the X-ray spectral shape of WZ Sge remains practically unchanged between the {\it XMM}-Newton and Swift observations.
    However, after correcting for inter-stellar
    absorption, the intrinsic luminosity is estimated to be ${\rm L^{Una}_X=(2.65\pm0.06)\times 10^{30}}$ erg s${\rm ^{-1}}$ and ${\rm L^{Una}_X=(1.57\pm0.03)\times 10^{30}}$ erg s${\rm ^{-1}}$ in 2003 and 2011,
    respectively. During the Swift/XRT observation, the
    observed flux is a factor
    $\simeq 2$ lower than that observed by {\it XMM}-Newton,
    but is similar to the quiescent levels observed various times before the
    2001 outburst.}
{}

}

   \authorrunning{Nucita et al.}
   \titlerunning{$XMM$-Newton and Swift observations of WZ Sge}
   \maketitle

\section{Introduction}

A cataclysmic variable (CV) is a binary system {consisting of a white
dwarf primary} which accretes matter from a low mass companion via Roche lobe overflow
(for a review, {see} \citealt{warner1995}). Systems with {a primary with} a relatively low
magnetic field ($\ut<$0.1 MG) are {expected to accrete} via a Keplerian disk. In {such a} case, half of the total
potential gravitational energy is dissipated by the viscosity, with the {remainder being} radiated away by the
boundary layer. {The} spectral energy distribution emitted by the accretion disk peaks
in the optical and ultraviolet bands, while the boundary layer radiates predominantly
in the extreme ultraviolet and X-rays. Typical X-ray luminosities of CVs are in the range
$10^{30}$--$10^{32}$ erg s$^{-1}$ (see e.g. \citealt{lamb82}, \citealt{baskill}, \citealt{erik}). {\it XMM}-Newton (\citealt{jansen2001})
is particularly {useful} for studying quiescent CVs as
its large effective area allows to detect faint sources
{in general and during} dips and eclipses {in particular}.
Moreover, the possibility to observe the source simultaneously in the optical {or ultraviolet (UV)} bands with the optical monitor (OM) opens the possibility to
study the correlations between light curves of the same source in different wavelengths {taken at exactly the same time}.

WZ Sagittae (hereafter WZ Sge) is currently known to be the closest CV ($43.5\pm0.3$ pc, see \citealt{harrison2004}).
It {reaches} $V\simeq 7-8$ {during outbursts} (e.g. \citealt{patterson2002}, \citealt{kuulkers2011}); it spends most of
the time, however, in a quiescent state characterized by rather modest optical magnitudes
in the range $14-16$ (e.g, \citealt{steeghs}, \citealt{kuulkers2011}). It has a {short orbital period of $\simeq 81.6$ min} (Krzemi\'nski 1962;
Warner 1976).
Apart from showing {a large outburst amplitude}, 
WZ Sge has also a long outburst recurrence time: it goes into outburst every 20-30
years. {In the literature,
there are reports of large outbursts of WZ Sge in 1913, 1946, 1978 and 2001 (see e.g.,
\citealt{mayall1946}, \citealt{brosh1979}, \citealt{mattei2001}, \citealt{godon2004}, \citealt{ishioka2001}
and references therein). For a the historical record of these
observations, we refer to \citet{kuulkers2011}.
Several observational campaigns were devoted to the study of the source
characteristics in detail during the 2001 outburst (see, e.g., }
\citealt{patterson2002}, \citealt{knigge2002}, \citealt{long2003}, \citealt{sion2003}).

{One prominent scenario for the long outburst recurrence time is
that the inner part of the accretion disk is truncated by the magnetic field
of the white dwarf (see, e.g., \citealt{warner1995}, \citealt{hameury1997}).
In this scenario, the $\simeq 28$ s periodic modulation in the optical
data (see, for example, \citealt{patterson1998}, \citealt{lasota1999}), is interpreted as
possibly related to the white dwarf spin period.  In Sect. 4, however, we discuss
a counterargument, suggesting that other scenarios should be considered.}

{The component masses of WZ Sge are still not well known.
The photometric solution of \citet{smak1993} gives a mass of the white dwarf of $M_{wd}\simeq 0.45$ M$_{\odot}$ and a mass ratio $q\simeq 0.13$, whereas
\citet{spruit1998}, who modeled the hot spot at which the mass stream transferred from the
companion hits the outer accretion disk, obtain
$M_{wd} \simeq 1.2$ M$_{\odot}$ and a mass ratio $q\simeq 0.075$. As shown by phase-resolved spectroscopy (\citealt{steeghs}), the binary system is
characterized by a primary white dwarf with mass in the range $0.88$ M$_{\odot}$--$1.53$ M$_{\odot}$ and
a low mass companion of $0.078$ M$_{\odot}$--$0.13$ M$_{\odot}$ which is close to the brown dwarf mass threshold. If the mean velocity of absorption lines
is interpreted as being due to gravitational red-shift (and one uses the mass-radius relation),
then the mass of the primary {is inferred} to be $(0.85\pm 0.04)$ M$_{\odot}$.
In the present work, we use the latter value for the mass of the white dwarf in
WZ Sge, i.e. $0.85$ M$_{\odot}$.} 

WZ Sge has been intensively observed in the X-ray band. \citet{patterson1998} described
both the ROSAT and ASCA observations (as well as Einstein and EXOSAT ones)
{obtained in quiescence} . This analysis was
successively re-done by \citet{gun2005} who reported a
quiescent 0.1--2.4 keV flux (as obtained from ROSAT PSPC in 1991) of $\simeq 2.8 \times 10^{-12}$  erg cm$^{-2}$ s${\rm ^{-1}}$ (corresponding to a
luminosity of $\simeq 6.3\times 10^{29}$  erg s${\rm ^{-1}}$ for a distance of $\simeq 43.5$ pc). In addition,
\citet{hasenkopf}, re-analyzing a 1996 ASCA observation of WZ Sge, found a 0.5--10 keV flux
of $\simeq 4.7\times 10^{-12}$  erg cm$^{-2}$ s${\rm ^{-1}}$, thus implying a
luminosity of $\simeq 1.0\times 10^{30}$  erg s${\rm ^{-1}}$.
Furthermore, the 2001 outburst of WZ Sge was observed in X-rays (see, {e.g.,}  \citealt{wheatley2001},
\citealt{kuulkers2002}, \citealt{wheatley2005}).

In this paper we present the result of $\sim 9.9$ ks {\it XMM}-Newton
and $\sim 1.4$ ks Swift observations of WZ Sge acquired
in 2003 and 2011, respectively, i.e. almost two and ten years after the {most recent} outburst.
{The {\it XMM}-Newton data were already reported by \citet{mukai2004},
who used a multi-temperature plasma to describe the observed WZ Sge X-ray spectra and
found a 2-10 keV band flux of $\simeq 7.0\times 10^{-12}$  erg cm${\rm ^{-2}}$ s${\rm ^{-1}}$
(i.e., much larger than the flux inferred by using the 1996 ASCA data).
We, here report on  
a coherent periodicity of $\simeq 28.96$ s in the same {\it XMM}-Newton observation, when 
using all the information down to $0.2$ keV. The detected periodicity is close 
to that found in the optical reported by \citealt{mukai2004}.}

The paper is structured as follows: in Sect. \ref{s:xmm1} and \ref{s:swift1} we present the available data
and give details about the {\it XMM}-Newton and Swift data reduction, respectively;
in Sect. \ref{s:result} (and related sub-sections) we present the results
of our timing and spectral analysis. Finally, in Sect. \ref{s:conclusion}
we conclude on our observations.

\section{Observations and data reduction}
\subsection{{\it XMM}-Newton}
\label{s:xmm1}
WZ Sge was observed by {\it XMM}-Newton
(observation ID 0150100101) for $\simeq 9.9$ ks
starting on $2003$ May 16 (14:52:0.7 UT).
The target was observed by the three {types of} X-ray instruments
(see, e.g., \citealt{jansen2001}): RGS 1 and 2, EPIC-MOS 1 and 2 operating in small window mode,
and EPIC-pn in full frame mode, and by the Optical Monitor (OM) on board
{\it XMM}-Newton. Here, we concentrate on the analysis of the data acquired by the MOS, pn
and OM cameras.

The observation raw data files (ODFs) were processed using the {\it XMM}-Science
Analysis System {(SAS, version $13.0.0$) and with up-to-date current calibration files (CCF)}. The data for the EPIC cameras
were processed by running the {\it emchain} and {\it epchain} tools, while the
{\it omfchain} pipeline was executed in order to obtain the optical (background corrected)
light curve of WZ Sge.
 \begin{figure*}[t]
\vspace{0.05cm}
\begin{center}
\epsfxsize=6.5in \epsfysize=5.5in \epsffile{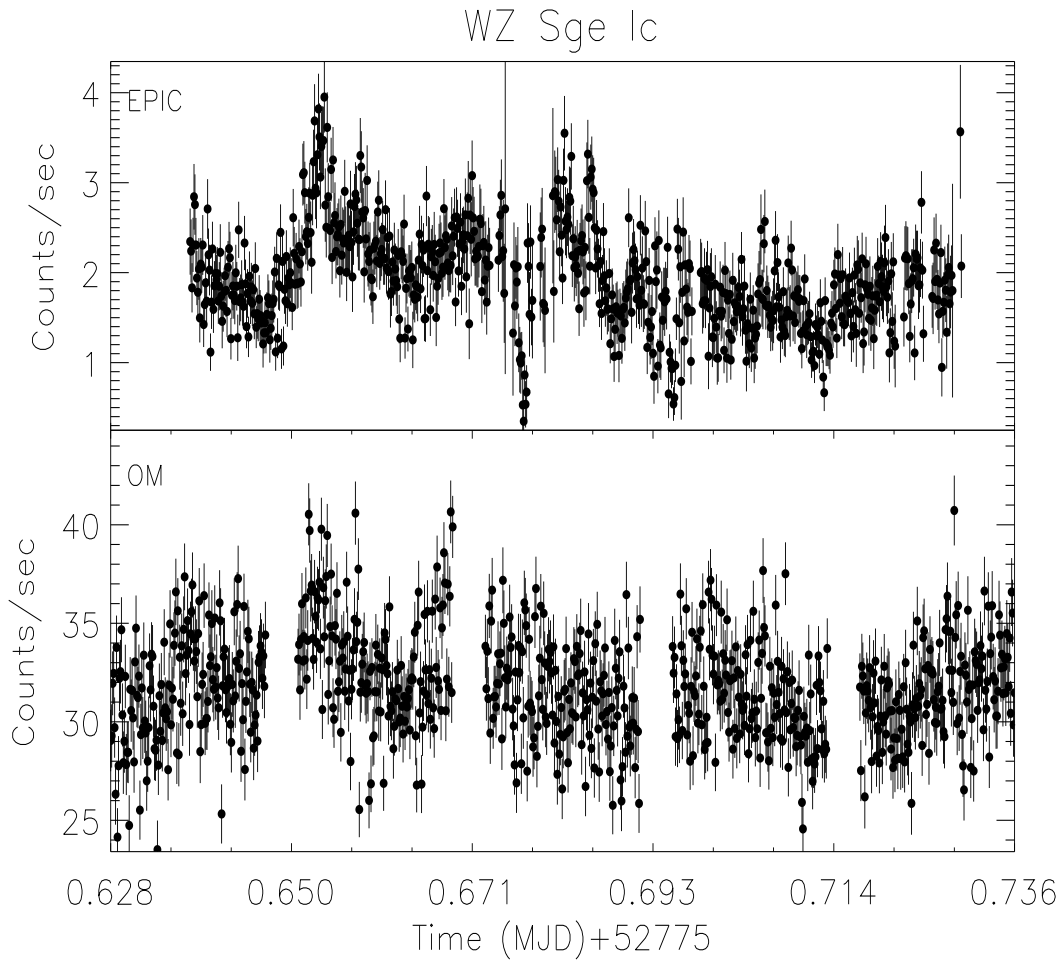}
\end{center}
\caption{Upper panel: $0.3$--$8$ keV {\it XMM}-Newton/EPIC (averaged) background- and barycentric corrected light curve of WZ Sge
with a time resolution of $10$ \,s. Bottom panel: OM barycentric corrected light curve with a bin size of $10$ \,s.}
\label{f1}
\end{figure*}
Following the standard screening procedure described in the \citet{xrps}, we extracted
light curves above 10 keV for the full MOS and pn cameras. Hence,
by identifying and discarding parts of the observation affected by high levels of background
activity, the effective observation
exposure times resulted in $\simeq 4.3$ ks and $\simeq 2.2$ ks for the MOS and pn cameras,
respectively. While the events collected during the
good time intervals were used in the spectral analysis, the timing analysis
was performed without applying any time filter as
the introduction of gaps may lead to artifacts.

The $X$-ray emission from the source was extracted from a circular region centered on
the nominal position of WZ Sge
and with a radius chosen
in order to contain at least $80\%$ of the total source energy. The background signal was extracted from circular regions on the same chip.
{Finally, the source light curves (one per each EPIC camera) were obtained after subtracting the background counts. We furthermore checked that the resulting data 
did not show residual effects due to the high energy background. We applied the Solar System barycenter correction, ensuring that 
the event times were in barycentric dynamical time instead of spacecraft time; the SAS task {\it epiclccorr} was used in order to account for absolute
and relative corrections.

Following the standard procedures described in the \citet{xrps}, we filtered the event list files of the MOS 1, MOS 2 and pn cameras
in order to exclude any background flare. {Here, we adopted the thresholds of $0.3$ count s$^{-1}$ and $0.4$ count s$^{-1}$ for MOS 1 (MOS 2) and pn, respectively.}
Then, we accumulated the spectra of the source from a circular region centered on
the nominal WZ Sge position, while the spectra of the background were extracted from nearby circular regions. Hence,
all the spectra were rebinned to ensure at least 25 counts per energy bin.

{The EPIC MOS 1, MOS 2, and pn source light curves were extracted over the energy 
range 0.3-8 keV with 10 s and 2 s binning, the former for plotting purposes and the latter for period searching. 
Finally, the three background-subtracted light curves were combined (averaged) in order to increase the signal-to-noise ratio.}

In addition, the fast mode of the OM telescope allowed us to obtain the {UV}
light curve (in the UVW1 filter, $200-400$ nm) of the source.
In the UVW1 filter,
the observed average source magnitude is $ 13.45\pm0.06$ which, for the standard Vega magnitude to flux conversion\footnote{More information
on the conversion between OM count rates to magnitude and fluxes is available at
\texttt{http://xmm.esac.esa.int/sas/current/watchout}.},
corresponds to a quiescent flux of $\simeq 1.57\times 10^{-14}$ erg cm$^{-2}$ s$^{-1}$ \AA$^{-1}$.}

\subsection{Swift/XRT}
\label{s:swift1}
{Swift/XRT observed the source on
three occasions in 2011 November, but we only used
the observation taken on 2011 November 14 ($10:00:00$ UT; ID 00032125002), which had the longest exposure ($\simeq 1.4$ ks)\footnote{The 
other observations, with IDs 00032125001 and 00032125004, have exposure times of only $\simeq 594$ s and
$\simeq 51$ s, respectively.}.
The Swift data were analyzed using standard procedures
(see \citealt{burrows}) and the latest calibration files available\footnote{The latest
Swift related softwares and calibration files can be found at
\texttt{http://heasarc.nasa.gov/docs7swift/analysis/}.}. In particular, we processed the XRT products
with the {\it xrtpipeline} (v.0.12.6) task, applied the standard screening criteria by using ftools (Heasoft v.6.13.0) and,
with the {\it xselect} task, we extracted the source spectra and light curves 
(in the 0.3-10 keV band) from a circular region (with radius of $\simeq 47\arcsec$)
centered on the target nominal coordinates. 

The background spectra and light curves were accumulated from a circular region with the same radius as the source extraction region. 
We first corrected the source light curve for losses caused by bad columns by using the {\it xrtlccorr} task and then we subtracted the background light curve  
(scaling for the extraction areas) by using the {\it lcmath} tool. 
We used the {\it xrtmkarf} task to generate the ancillary response files and account for different extraction areas of the source and background,
vignetting and PSF corrections.
}


\section{Analysis and results}
\label{s:result}
\subsection{{\it XMM}-Newton temporal analysis results}
The MOS 1, MOS 2 and pn single light curves have an average count rates of $1.13 \pm 0.87$ count s$^{-1}$,
$1.10 \pm 0.86$ count s$^{-1}$, and $3.53 \pm 1.56$ count s$^{-1}$, respectively, while the combined (averaged) background corrected
($0.3$--$8$ keV) light curve has an average count rate of $1.92 \pm 0.69$ count s$^{-1}$ {per instrument}.
The combined (MOS and pn averaged) {\it XMM}-Newton/EPIC light curve is {shown} in Figure \ref{f1} (upper panel). 
For comparison, we also {present} (bottom panel) the OM (UVW1 filter) light curve of WZ Sge binned at 10 s.

\begin{figure*}[t]
\vspace{0.05cm}
\begin{center}
\epsfxsize=6.5in \epsfysize=5.5in \epsffile{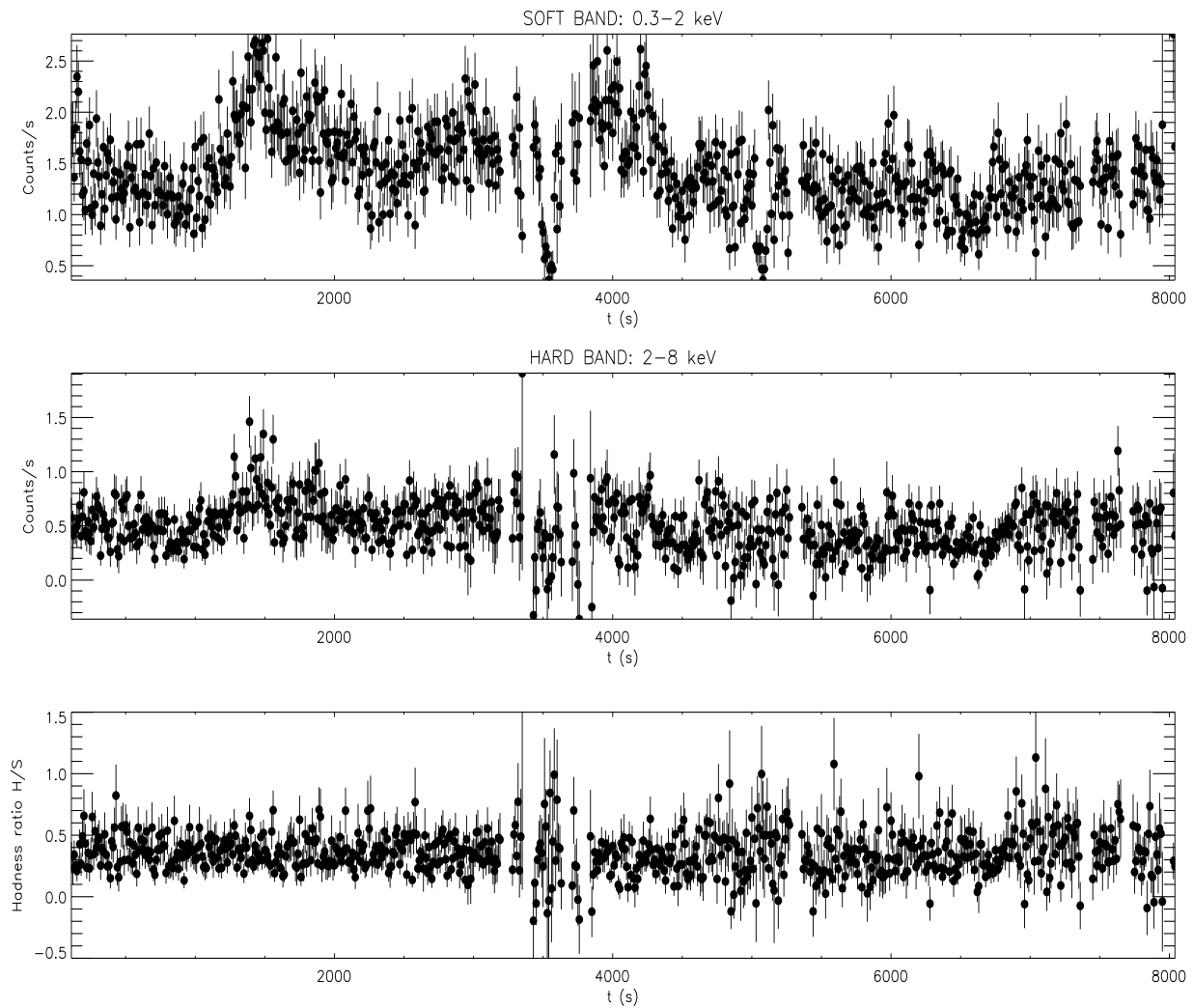}
\end{center}
\caption{Upper and middle panels: the light curves (10 s bins) in the soft (0.3-2 keV) and hard (2-8 keV) bands. Bottom panel: the hardness ratio.
For clarity, the time axis is given in seconds with time zero corresponding to the start of the observation.}
\label{f2}
\end{figure*}

The source is variable on several time-scales and shows
structures that may resemble dip features produced by absorbing matter intervening
along the line of sight (see Sect. 4).
To test this hypothesis,
we produced background-corrected light curves in the energy ranges
$0.3-2.0$ keV and $2.0-8.0$ keV (soft and hard bands\footnote{
The MOS 1, MOS 2 and pn single light curves in the soft band have average count rates of $0.80 \pm 0.14$ count s$^{-1}$,
$0.79 \pm 0.13$ count s$^{-1}$, and $2.68\pm 0.61$ count s$^{-1}$, respectively. The combined (averaged) $0.3$--$2$ keV light curve
has a typical count rate of $1.43 \pm 0.26$ count s$^{-1}$. When the hard band is considered, we find $0.33 \pm 0.27$ count s$^{-1}$,
$0.31 \pm 0.25$ count s$^{-1}$, and $0.84\pm 0.39$ count s$^{-1}$ for the MOS 1, MOS 2 and pn, respectively, while the combined hard light curve
has an average count rate of $0.49\pm 0.18$ count s$^{-1}$.}, respectively) and
define the hardness ratio as the ratio between the hard light curve to the soft one. The result of this approach is given in
Figure \ref{f2}: the hard and soft light curves are given in the upper and middle panels, respectively, while the hardness ratio
is shown in the bottom panel.
Inspection of this figure shows that the hardness ratio remains practically constant during the observation, indicating
that the photon counts in the soft and hard energy ranges strongly correlate. 

\begin{figure*}[t]
\vspace{0.05cm}
\begin{center}
\epsfxsize=6.5in \epsfysize=5.5in \epsffile{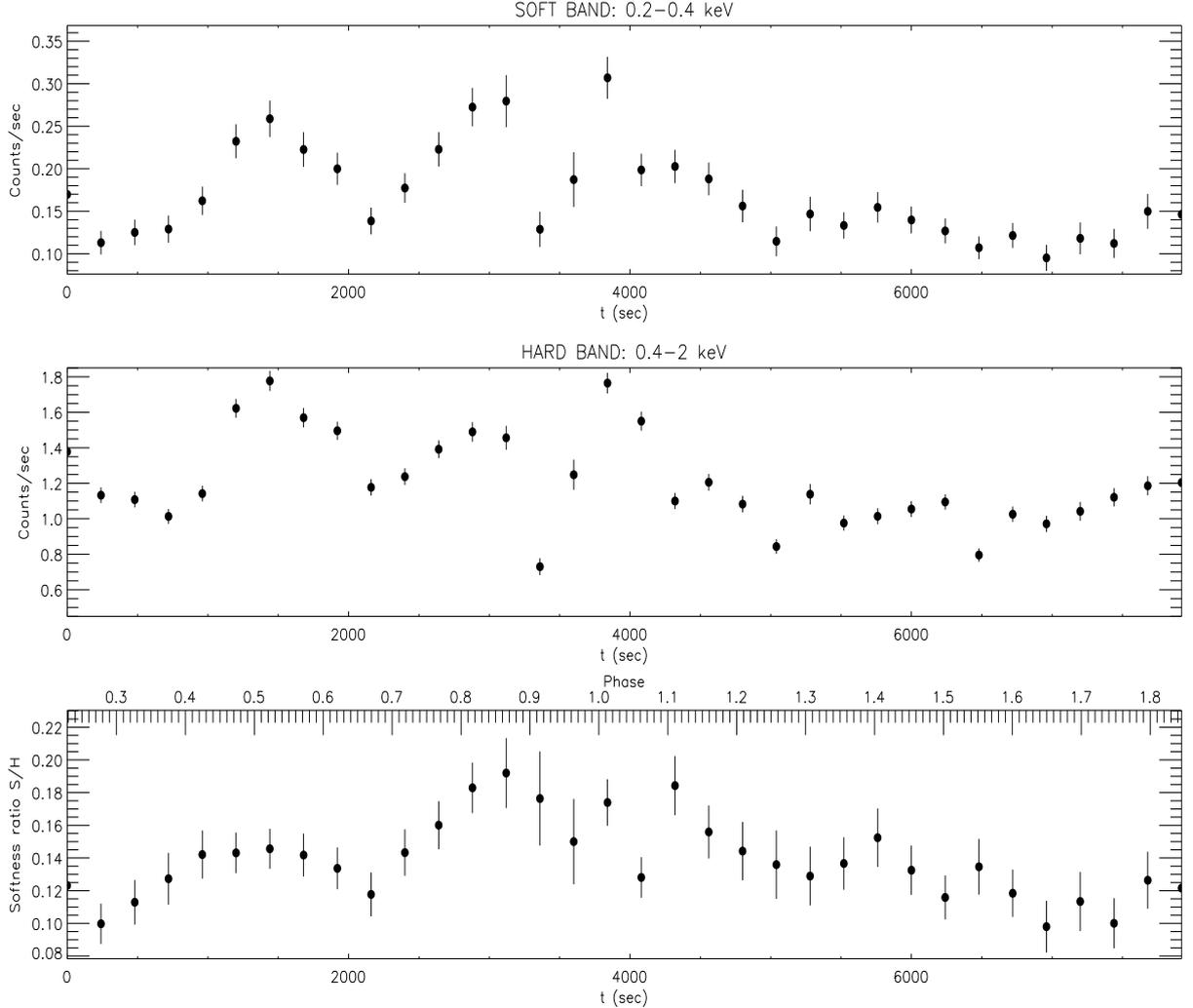}
\end{center}
\caption{The WZ Sge light curves between 0.2 keV and 0.4 keV (upper panel) and between 0.4 and 2 keV (middle panel). The bottom panel shows the softness ratio
(see text for details).}
\label{softnessratio}
\end{figure*}

When analyzing the 1991 ROSAT PSPC data of WZ Sge in the energy band up to 2 keV, by averaging over various orbital cycles, \citet{patterson1998} 
found that the soft to hard X-ray ratio 
(also known as {\it softness ratio}, in the energy bands 0.2 keV - 0.4 keV  and 0.4 keV - 2 keV) 
clearly showed a modulation at the orbital period, as well as a dip at orbital phase $\simeq 0.7$.
{Although, we do not have the repeated phase coverage to study the average properties of the dips, 
we extracted the light curves 
with a bin size of $240$ s corresponding to about 0.05 orbital cycle.} The resulting light curves are shown in Figure \ref{softnessratio} (upper panels) together with the softness
ratio plotted in bottom panel. Here, for convenience, we give the horizontal axes in seconds, and in orbital phase by
using the ephemeris given in \citet{patterson1998}. The softness ratio has a shape similar to that already observed in ROSAT PSPC data.
In particular the curve shows a modulation at the orbital period and, although not dramatic, a dip appears close to the orbital phase $\simeq 0.7$. 

\begin{figure}[htbp]
\vspace{8.0cm} \includegraphics{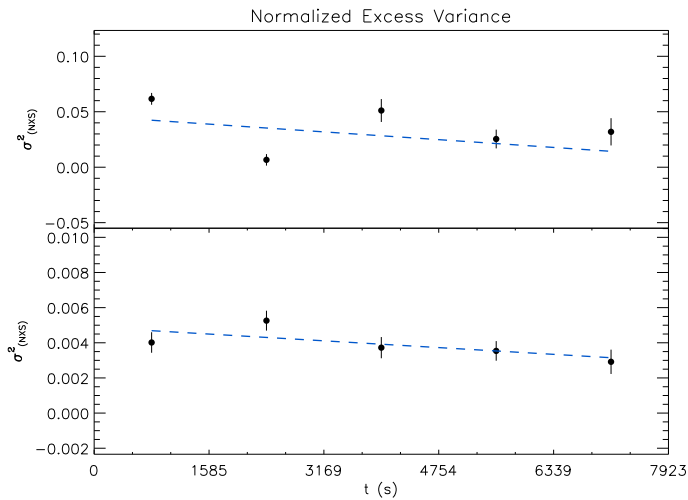}
\caption{The excess variance, $\sigma^2_{NXS}$, of the combined EPIC (upper panel) and OM (bottom panel) light curves of WZ Sge (see text for details).}
\label{f6}
\end{figure}

We quantified the variability amplitude of the high-energy and optical light curves, using the normalized excess
variance ($\sigma^2_{NXS}$, see e.g.
\citealt{nandra1997} and \citealt{edelson2002}) and evaluated the associated errors according to eq. (11) in \citet{vaughan2003}.
{The result of our analysis is shown in Figure \ref{f6}. Here, we give the excess variance for the combined EPIC (upper panel)
and OM (bottom panel) light curves (each with a bin size of 10 s) calculated in 5 (equally spaced)
time bins}. Since negative values of $\sigma^2_{NXS}$ indicate absence or very small variability in the time series, we conclude that both the
high energy and optical data of WZ Sge present a certain degree of intrinsic variability, which seems to vary with time.
In fact, when we fitted (separately) each normalized excess variance with a linear function
(represented in both panels of Figure \ref{f6} by the dashed lines), we found that the rate of change in time of $\sigma^2_{NXS}$ for the $X$-ray and optical data
is  $\simeq -4.4\times 10^{-6}$ and $\simeq -2.4\times 10^{-7}$, respectively.

\begin{figure}[htbp]
\vspace{6.7cm} \includegraphics{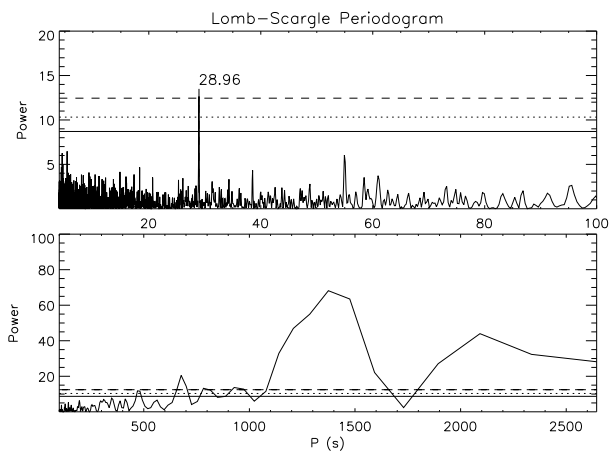}
\caption{The Lomb-Scargle periodogram of the WZ Sge combined EPIC light curve in the 0.3-8 keV band (see text for details). }
\label{f4}
\end{figure}

We blindly searched for periodicities in the time range 2 s - a few hours in the $X$-ray light curve by applying the
Lomb-Scargle technique (\citealt{lomb,scargle1982}). In particular,
we used $\nu_{min}=1/(3 T_{obs})$ and $\nu_{max}=1/(2 \delta t)$, {with $T_{obs}$ the duration
of the observation and $\delta t$ the associated time step, as the minimum and maximum values of the frequency range to be searched for periodicity.
Note that by using the minimum frequency $\nu_{min}$ we implicitly require at least three full cycles per observational window. The analysis
resulted in the periodogram shown in Figure \ref{f4}. 
In the upper panel, we show the periodogram in the period range 10-100 s while, in the bottom panel, we show the periodogram in the range 100-2600\,s.
The significance of each peak appearing in the periodogram was evaluated by following the recipe described by
\citet{lomb,scargle1982}. In particular, we compared the height of each peak with the power threshold corresponding to a given
false alarm probability in white noise simulations:  the three horizontal lines given in Figure \ref{f4} correspond
to false alarm probability thresholds of $68\%$ (solid line), $90\%$ (dotted line), and $99\%$ (dashed line), respectively.
}

\begin{figure}[htbp]
\vspace{8.0cm} \includegraphics{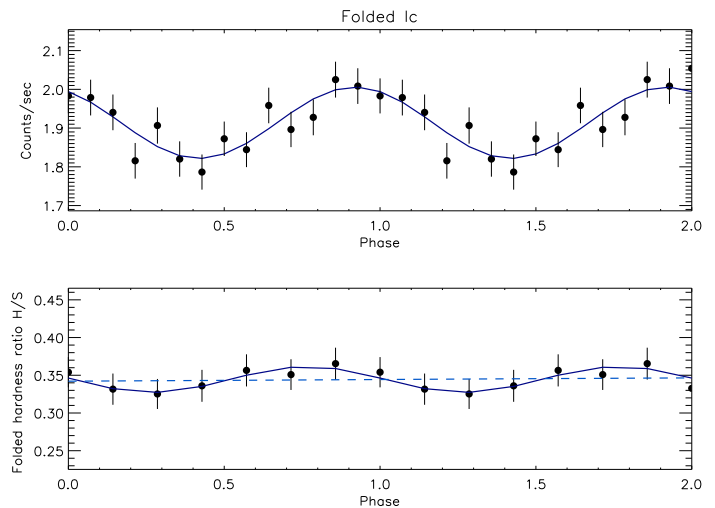}
\caption{The high-energy light curve of WZ Sge (0.3-8 keV; upper panel) and the hardness ratio (lower panel) folded at $\simeq 28.96$\,s with 15 bins and 8 bins per cycle, respectively. 
See text for details.}
\label{f5}
\end{figure}
It is clear that we detect a periodicity of $28.96^{+0.02}_{-0.01}$\,s in the 0.3-8 keV light curve (Figure 5, upper panel), being 
the $1 \sigma$ error on the detected
period {estimated with the technique described in \citet{carpano2007}}. 
This period was confirmed (within the quoted error range) by using the epoch folding method. In particular,
we iteratively folded the light curve at a trial period $P$, fitted the resulting light curve with a sine function and searched for the period that
minimized the $\chi^2$ statistics. 
The period is consistent with {the low significance signal reported by \citet{mukai2004}} when analyzing the same set of {\it XMM}-Newton data in the 2-10 keV energy band,
as well as with the coherent periodicity detected in 2003 in the MDM 2.4 m telescope data by the same authors. 
We confirm, in accordance with \citet{mukai2004}, that the $28.96$\,s peak becomes less significant when we consider the light curve
in the 2-10 keV band. Moreover, {we do not detect any coherent period at $\simeq 27.87$ s}, i.e. the {presumed} spin period of the white dwarf (see Sect. 1).

{In the bottom panel of Figure \ref{f4}, we show the Lomb-Scargle periodogram in the period range 100-2600\,s.
Note that the features appearing at $\simeq 1400$ might be associated to longer time-scales present in the light curves.

We folded the EPIC light curve and the hardness ratio curve on the 28.96 s period with 15 bins and 8 bins per cycle, respectively.
The result is shown in Figure \ref{f5}. The solid line in the upper panel corresponds to a sinusoidal fit\footnote{We performed the sinusoidal fit 
with the function $s(t)=A\sin(2\pi \phi +B)+C$, where $\phi$ is the phase and $A$, $B$ and $C$ the free parameters.} to the data 
having $\chi^2=0.8$ for 26 d.o.f.: in particular, the amplitude of the sine signal results to be $A=0.09\pm 0.01$ count s$^{-1}$. 
In the bottom panel, the solid and dashed lines represent a sinusoidal (with amplitude $A=0.017\pm 0.003$  and $\chi^2=0.14$ for 12 d.o.f.) 
and linear (with constant value of $C=0.34\pm0.01$ with $\chi^2=0.5$ for 13 d.o.f.) fits to the folded hardness ratio. 
As it is evident, the hardness ratio data is consistent with being constant in time although it appears to show a low modulation by eye.
\begin{figure}[htbp]
\vspace{6.7cm} \includegraphics{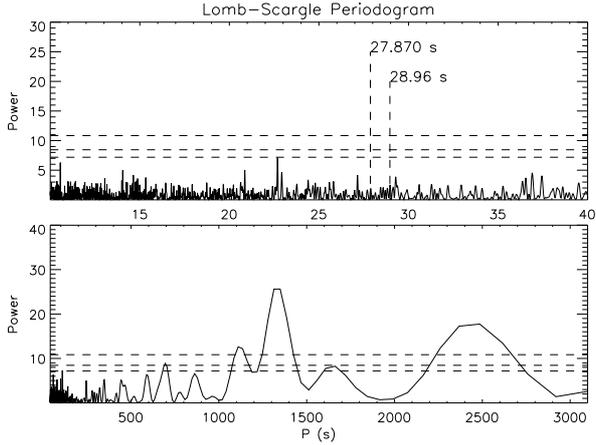}
\caption{The same as in Figure \ref{f4} for the OM data of WZ Sge. }
\label{f4bis}
\end{figure}

We applied the same procedure to the OM light curve (see Figure \ref{f4bis}). We do not detect the
coherent period at $\simeq 28.96$ s {nor the periodicity of $\simeq 27.87$ s}. We also verified with a phase resolved
periodogram that the peak appearing at $\simeq 22$ s is not significant. }

\subsection{Swift/XRT temporal analysis results}
\label{s:result2}
\begin{figure}[htbp]
\vspace{7.5cm} \includegraphics{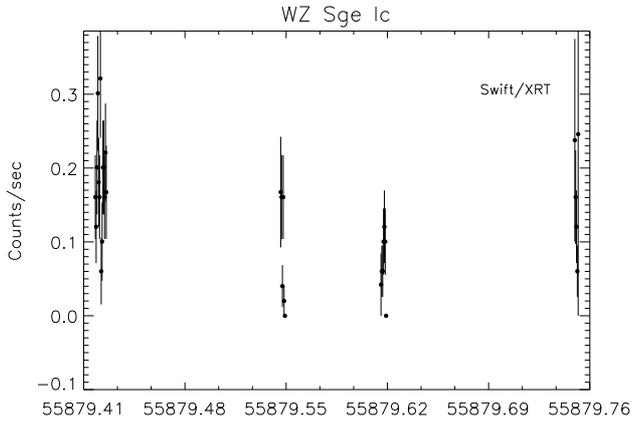}
\caption{The Swift/XRT light curve of WZ Sge in the 0.3-10 keV energy band with a time bin size of 50 s.}
\label{f7}
\end{figure}

{The quality of the 2011 Swift/XRT data do not allow to perform a
detailed timing analysis similar to the $XMM$-Newton data set: the observed Swift/XRT average count rate is $\simeq 0.14$ count s$^{-1}$ 
during $\simeq 1.4$ ks. We obtained the quiescent WZ Sge light curve in the 0.3-10 keV energy band with a bin size of 50 s, which is shown 
in Figure \ref{f7}. Over the course of the observation we do not see significant variability.}

\subsection{{\it XMM}-Newton and Swift/XRT X-ray spectral analysis}
\label{s:result3}

The {\it XMM}-Newton and Swift/XRT source spectra
(including the response matrices, ancillary files and background spectra)
were {read} into the XSPEC package (version 12.4.0) for the spectral
analysis and fitting procedure. A first fit attempt showed that the
spectral shape of the source did not change significantly between the times of the {\it XMM}-Newton and Swift/XRT
observations. Hence, we fitted the data with all the model
parameters linked together, apart from a multiplicative dimensionless constant,
which can take different values
for the {\it XMM}-Newton and Swift/XRT spectra. The multiplicative factor 
{mostly accounts for any flux change of the source that might have happened between the two observations.}

We tried, without success, to fit the spectra by using a single thermal plasma component absorbed by neutral gas ({\rm Mekal} and {\rm Phabs} in XSPEC).
Fixing the hydrogen column density to the average value\footnote{See the on-line calculator available at 
\texttt{http://heasarc.gsfc.nasa.gov/cgi-bin/Tools/w3nh/w3nh.pl} which gives the integrated neutral 
hydrogen column density through the Galaxy.} found in the direction of the
target (${\rm n_H\simeq 2\times 10^{21}}$ cm${\rm ^{-2}}$, \citealt{dickey}) 
and/or allowing the metallicity abundance to vary did not improve significantly our fit. In particular, we noted the existence of residuals
at low energies around the iron L-shell complex at $\simeq 1$ keV. 
These kind of residuals may be due to two effects: the improper modeling of photo-electric absorption and the
fluorescence from cold material together with a multi-temperature structure of the spectrum (see e.g. \citealt{baskill}).
In order to account for such a line,
we added an extra Gaussian component ({\rm Gauss} in XSPEC). 
 \begin{figure}[htbp]
 \vspace{8.cm} \includegraphics{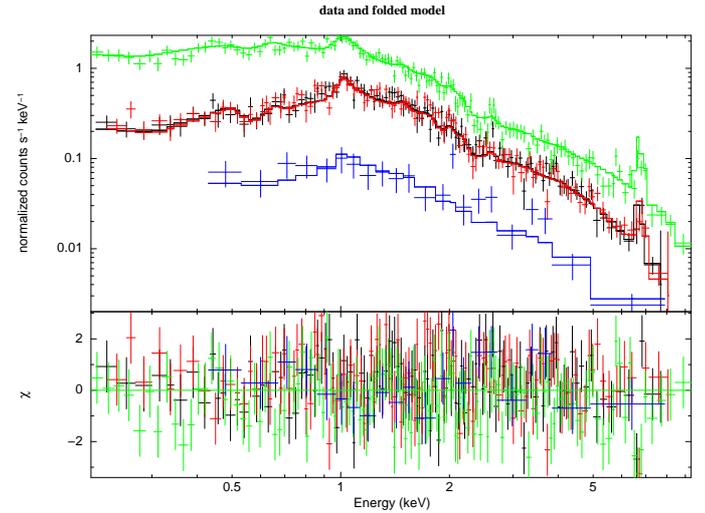}
 \caption{Upper panel: the MOS (black and red crosses), pn (green crosses) and XRT (blue crosses) spectra together with the best-fit model (solid lines). Bottom panel: 
the residuals between data and the best-fit model.}
 \label{f8}
 \end{figure}
Thus, our model, {\rm K*phabs(mekal + gaussian)}, 
consists of seven free parameters, i.e. the hydrogen column density $n_H$ towards the source, the plasma temperature $kT$ and normalization $N$
of the emission model
the position $E$, width $\sigma_E$, and normalization $N_E$ of the Gaussian line, and the multiplicative constant $K$ which accounts
for any difference in flux among the spectra from the different instruments. 
We fixed $K$ to 1 for the {\it XMM}-Newton/MOS 1, MOS 2, and pn spectra while we allowed it to vary for the Swift/XRT data.
The other parameters of the {\it Mekal} model, i.e. the solar abundance and hydrogen number density, were fixed to the respective
default values.

{Using the above model, we obtained $N_H=(0.031\pm0.004)\times 10^{22}$ cm$^{-2}$,
$KT=6.9\pm0.5$ keV, $N=(5.5\pm 0.1)\times 10^{-3}$, $E=1.01_{-0.02}^{+0.01}$ keV,  $\sigma_E\ut < 0.07$ keV,
$N_E= (9.6_{-2.0}^{+3.0})\times 10^{-5}$ (with $\chi^2=1.24$ for 393 d.o.f.). 
Based on the multiplicative factor, we find that the Swift/XRT data show an overall decrease in flux by a factor $K=0.59\pm 0.07$
as determined by our best fitting procedure.

The Gaussian line at $\simeq 1$ keV may be due to the L-shell complex or to the H-like line of Ne expected around this energy. Hence, we tried to fit the data with a more physical model 
consisting of a two-temperature plasma model. The best fit procedure resulted ($\chi^2=1.16$ for 394 d.o.f.) in a hydrogen column density formally 
consistent with the value quoted above, and plasma temperatures of $KT_1=9.04^{+1.15}_{-0.91}$ keV and $KT_2=1.31^{+0.08}_{-0.13}$ keV with normalizations 
$N_1=(5.0\pm 0.1)\times 10^{-3}$ and $N_2=(4.9\pm 0.1)\times 10^{-3}$, respectively. Also in this case, the Swift/XRT data show a flux decrease by a factor $K=0.59\pm 0.07$.

{In Figure \ref{f8}, we present the MOS 1 (black), MOS 2 (red), pn (green) and XRT (blue) spectral data in
the energy band 0.2-10.0 keV together with the two-temperature best-fit model (solid lines).}
The total absorbed flux in the 0.2-10.0 keV band is ${\rm F^{Abs}_{0.2-10.0}=(1.09\pm0.02)\times 10^{-11}}$ erg s$^{-1}$ cm${\rm ^{-2}}$ for the
{\it XMM}-Newton 2003
observation (i.e. consistent with what found by \citealt{mukai2004}) and ${\rm F^{Abs}_{0.2-10.0}=(6.47_{-0.74}^{+0.69})\times 10^{-12}}$ 
erg s$^{-1}$ cm${\rm ^{-2}}$ for the Swift 2011 observation. The errors quoted are at the $90\%$
confidence level. As noted by \citet{mukai2004}, the 2003 WZ Sge flux (restricted to the 2-10 keV energy band)
is higher than the 1996 May ASCA data ($\simeq 2.9\times 10^{-12}$ erg s$^{-1}$ cm${\rm ^{-2}}$).
The 2011 Swift data show that the high-energy signal returned to a similar level to that observed in
1996 which preceded the last source outburst.

Assuming a distance of $43.5\pm0.3$ pc (see, e.g., \citealt{harrison2004}), the X-ray luminosity of WZ Sge 
in 2003 was ${\rm L^{Abs}_X=(2.47\pm0.06)\times 10^{30}}$ erg s${\rm ^{-1}}$, while in 2011 it was ${\rm L^{Abs}_X=(1.41^{+0.17}_{-0.16})\times 10^{30}}$ erg s${\rm ^{-1}}$. The major contribution to the error in the luminosity comes from
the error associated to the source distance. 
Correcting for the absorption, we get an intrinsic luminosity of
${\rm L^{Una}_X=(2.65\pm0.06)\times 10^{30}}$ erg s${\rm ^{-1}}$ and ${\rm L^{Una}_X=(1.57^{+0.18}_{-0.17})\times 10^{30}}$ erg s${\rm ^{-1}}$ 
in 2003 and 2011, respectively.
}

\section{Discussion}
\label{s:conclusion}
\subsection{Quiescent X-rays}
In this paper, we presented the analysis of an archival {\it XMM}-Newton observation in 2003 (for a preliminar study see \citealt{mukai2004})
and newly acquired Swift data in 2011 of WZ Sge.

{WZ Sge's X-ray spectral properties in the 0.2-10 keV energy band 
remained practically unchanged between the 2003 and 2011 observations. 
We estimated an unabsorbed intrinsic luminosity of ${\rm L^{Una}_X=(2.65\pm0.06)\times 10^{30}}$ erg s${\rm ^{-1}}$ and 
${\rm L^{Una}_X=(1.57\pm0.03)\times 10^{30}}$ erg s${\rm ^{-1}}$ for the
2003 and 2011 observations, respectively. The luminosity in 2011 is a factor
$\simeq 2$ lower than that in 2003, indicating that WZ Sge returned to a level similar to
that observed prior to the last source outburst in 2001.

The high-energy light curves confirm
the existence of a dip close to the orbital phase $\simeq 0.7$. Although this feature is not strong, it also appears 
in the softness ratio light curve, similar to that seen using ROSAT/PSPC data
(\citealt{patterson1998}).
Dip structures in the light curves are naturally explained in the framework 
depicted by \citet{frankkinglasota1987} (see also \citet{smak1971}) which is
supported by the numerical simulation of \citet{hirose1991} and \citet{armitage1998}.
The model found its application in explaining periodic orbital dip features in the high-energy light curves of nova-like systems (see e.g. \citealt{hoard2010} and
\citealt{nucita2011}) and of magnetic white dwarfs (see \citealt{ramsay2009}).
According to this model, once the mass flow reaches the inferior conjunction
at the orbital phase 0.7, part of the accreting matter sets sufficiently high above the disk, thus obscuring the white dwarf and
producing the observed dip. 

\subsection{Periodicities at 27.87 and 28.96 s}

With an improved analysis (using all the available data down to $0.2$ keV) we find a coherent periodicity of $\simeq 28.96$ s in the 
2003 observation. This confirms the weak detection reported by \citet{mukai2004}; the period is close to that found in optical data 
reported by the same authors.
We did not detect the $27.87$ s oscillation attributed to the white dwarf spin (see e.g. \citealt{patterson1998}, \citealt{lasota1999}),
similar to \citet{mukai2004}. 
}

The origin of the 27.87 s {\sl and\/} the 28.96 s periods in WZ Sge has
been a long-standing puzzle.  For example, \citealt{robinson1978}
interpreted these two, distinct, periods as due to non-radial pulsations.
\citet{patterson1998} cautiously argued that
the 27.87 s period seen in the ASCA X-ray data was the spin period of
a magnetic white dwarf. \citet{welsh2003} presented a balanced review
of the two models, and pointed out the difficulties with both.  Given that 10
years have elapsed since then, during which a large body of recent
observations of non-radial pulsations in other low-accretion rate dwarf
novae have been obtained, we present a re-assessment of the models.

\citet{lasota1999} proposed that the 27.87 s is the white
dwarf spin period, while the 28.96 s signal is due to reprocessing of the
spin signal by a blob at the outer rim of the Keplerian disk.  While this
explanation is viable for an {\em optical} modulation at the 28.96 s period, it fails
to explain the 28.96 s {\em X-ray} period.
While intermediate polars often show
X-ray spin and sideband signals simultaneously (\citealt{norton}),
this is believed to be due to stream overflow -- mass
transfer stream that skirts the surface of the disk and is directly
captured by the magnetic field of the white dwarf.  It is hard to see
how the white dwarf can accrete directly from a blob at the {\sl outer\/}
edge of the disk.  If, instead, the Keplerian period at the inner edge of
the disk is 733.5 s (see, however, objections to this idea by \citealt{lasota1999}),
this could in principle lead to an X-ray modulation at the
28.96 speriod.  In this case, however, it would be difficult to avoid a
strong X-ray modulation at the 733.5 s period (\citealt{wynn1992}), given
the high inclination of the WZ Sge system.  That is, when the blob that
feeds the magnetic pole is on the Earth side of the white dwarf, the pole
that is facing the Earth would accrete more favourably.  This is likely to
lead to a higher observed X-ray flux than when the blob is on the far side.
In addition, both the inner and outer radii of an accretion disk are
not constant when accretion rate varies; it is not clear how a blob with
a Keplerian period of 733.5 s is always favoured, when both the optical
(\citealt{kuulkers2011}) and the X-ray (this work) brightness show secular
variability.

Further arguments against an intermediate polar model come from the hardness curves. 
We do not find any evidence for X-ray spectral variations along the 28.96 s
signal. Usually, in intermediate polars the $X$-ray emission is softer when brighter. 
We note that, indeed, WZ Sge is not classified as standard member of this class of objects (see e.g. \citealt{knigge2002}).

The above described weaknesses, however, may not be fatal for the magnetic CV model of the twin
periods. Nevertheless, the $XMM$-Newton detection of the 28.96 s signal
makes the argument that the 27.87 s period is the spin period of a
magnetic white dwarf somewhat weaker.

There is little doubt that the short period variability seen in another faint CV, GW Lib (see also below),
is due to non-radial g-mode pulsations of the white dwarf that dominates
its optical light in quiescence  (\citealt{vanzyl2004}).  Since then,
similar pulsations have been discovered in about a dozen of other faint,
white dwarf-dominated CVs (see, e.g., \citealt{szkody2010} and references
therein).  In the case of these accreting white dwarfs which are rapidly
rotating and have peculiar abundances, these pulsations are more complicated
than in the non-accreting ZZ Cet stars. For example, CV primaries may show
pulsations outside the ZZ Ceti instability strip.  The $<$30 s
periods in WZ Sge, however, are significantly shorter than those seen in GW Lib
type CVs ($>$200 s). Moreover, X-rays are generated by accretion, and
it is not clear how non-radial pulsations would modulate the X-ray flux.

In summary, the origin of the twin pulsations is as mysterious as ever.
The long-term stability of the intermittent 27.87 s period remains a strongest argument
for this to be the spin period of the white dwarf, but this leaves us
without a clear understanding of the 28.96\,s period.  It is interesting to
note, that \citet{Mukadam2013} found several puzzling features in the
pulsational variability of another CV, EQ Lyn.  In addition to possible ways to
reconcile these observations with our understanding of g-mode pulsations,
they considered alternatives models: r-mode pulsations and accretion disk
pulsations. We should maybe keep in mind such alternative possibilities when
considering WZ Sge.

\subsection{On the quiescent rate of accretion}

In addition to the twin periods of 27.87 s and 28.97 s, WZ Sge possesses
several characteristics that made it stand out among dwarf novae.  These
include the short orbital period, the quiescent spectrum which is dominated
by the white dwarf photosphere, the large outburst amplitude and the long
inter-outburst interval.  However, recent advances show that CVs with many
of these latter characteristics are in fact quite common. In particular, the Sloan
survey has revealed a large population of CVs near the period minimum
(P$<$ 88 min) whose spectra are often dominated by the white dwarf
photosphere (\citealt{Gaensicke}).  The {earlier} surveys did not go deep
enough to show the prevalence of this population. Many of these newly
discovered systems are candidate WZ Sge stars in terms of their outburst
characteristics -- they are generally seen in a quiescent dwarf nova-like
state since their discovery, so any outbursts must be infrequent.

The best studied such system is the aforementioned CV, GW Lib, whose discovery in fact predated
the Sloan survey.  Its well-documented 2007 outburst (\citealt{Byckling};
\citealt{Vican}) is the second known after the discovery outburst in 1983.
It has a 76.8 {min} orbital period, its quiescent spectrum is dominated by the
white dwarf photosphere, the outburst amplitude is large ($\sim$9 mag), and its
duration long ($\sim$26 day).  Surely, GW Lib presents a similar challenge
to the disk instability model that WZ Sge does.  Yet, despite intensive
observations motivated by its status as the prototype CV with non-radial
pulsations, no spin-period signature has ever been observed in GW Lib.

Of the many systems that share various degrees of similarity with WZ Sge
(\citealt{Gaensicke}), only V455 And (HS 2331+3905; \citealt{Araujo})
is known to be magnetic.  Intensive searches for additional
non-radial pulsators have not led to discoveries of magnetic CV signatures
among other WZ Sge-like systems. Unless all systems near the period minimum are sufficiently magnetic to create a hole in the disc,
yet somehow manage to hide any spin signatures, we must seek an explanation
for the long interval, long duration and
large amplitude outbursts that do not rely on the primary's magnetic field.
In particular, if the correlation found by \citet{patterson2011} between the
outburst recurrence time and the mass ratio is confirmed, some factor
directly related to the mass ratio is strongly implicated as the cause
of the long recurrence time in WZ Sge type systems; the magnetic field
of the white dwarf would be a second parameter, not the primary.

If that is the case, the twin periods of WZ Sge, whatever their origin,
may well be red herring in terms of understanding the outburst properties
of WZ Sge.  For example, while the detailed propeller model of
\citet{matthews2007} can still explain WZ Sge, it fails to explain
GW Lib whose outburst properties are similar to those of WZ Sge.
On a possibly related note, while the X-ray luminosity of WZ Sge
is low compared to dwarf novae with frequent outbursts (U Gem and SU UMa
types; \citealt{Byckling2010}), it is higher than that of GW Lib or the
Sloan-selected systems studied by \citet{reis2013}.  Given this, future
studies should strive to understand why the quiescent accretion rate in WZ Sge
is high compared to other WZ Sge systems, not why it is lower than in normal
dwarf novae.

\begin{acknowledgements}
KM thanks Paula Szkody for informative discussion on the latest results
of pulsations in low accretion rate CVs. This paper is based on observations from $XMM$-Newton, an
ESA science mission with instruments and contributions directly funded by ESA
Member States and NASA. Part of this work is based on archival data, software or on-line services provided
by the ASI Science Data Center (ASDC), Italy. We are also in debt with the anonymous Referee for pointing us
a problem in the Swift analysis. 
\end{acknowledgements}


\end{document}